# Tailoring "photonic hook" from Janus dielectric microbar

Yury E. Geints[1,*], Igor V. Minin[2,3,**] and Oleg V. Minin[2,3]

[1]V.E. Zuev Institute of Atmospheric Optics SB RAS, 1 Zuev square, 634021 Tomsk, Russia
[2]Tomsk State Politechnical University, Tomsk, 36 Lenin Avenue, 634050, Russia.
[3]Tomsk State University, Tomsk, 30 Lenin Avenue, 634050, Russia.
*ygeints@iao.ru; **prof.minin@gmail.com

**Abstract**

A new type of curved subwavelength light beam, a photonic hook (PH), emerged from the diffraction of a light wave at a mesoscale dielectric object with broken internal symmetry (Janus particle) was recently reported. For efficient and technically simpler generation of PHs, we propose a new type of asymmetric Janus microparticles, the orthogonal microbars, composed from materials with different refractive indices. Based on the numerical simulations by means of the Finite Elements Method, we investigate the physical mechanism of PH formation by analyzing the field intensity distribution and the energy fluxes near such particles. The influence of dielectric substrate is also studied for the first time. We show that by changing the refractive index contrast between two parts of Janus bar the shape and curvature of the PHs can be efficiently controlled. The criteria of PH quality for complex characterizing the photonic hook curvature is introduced and discussed as well.

## 1. INTRODUCTION

Transparent dielectric microobjects of various geometric shapes, physical properties, and interior composition are currently the focus of researchers' attention all over the world [1-9]. Such mesoscale microobjects, i.e., the objects with the characteristic spatial scales on the order of light wavelength, are fairly promising in obtaining light localization within extremely small area. Diffraction of electromagnetic radiation at microparticles produces highly-localized regions of enhanced intensity near their surfaces, the so-called photonic nanojets (PNJs) [1,2,7,9]. The physical nature of spherical particle aided nanojets formation is related to the aberrational character of optical near-field focusing at a transparent wavelength-sized particle [7,11]. Under these conditions the formation of a focal region with the spatial super-resolution (up to subdiffraction in size), high intensity, and increased length is possible due to the constructive interference between the scattered and incident fields in the particle shadow.

As known, the key parameters of photonic jet depend on the refractive index contrast, particle geometry and spatial dimensions [1-4,7,9]. Recently, a new type of subwavelength-scaled curved beams – a "photonic hook" (PH), was proposed [12] based on PNJ phenomenon [1,2,7]. To demonstrate the main principle of PH formation and its properties, a Janus particle in the form of a cuboid with broken symmetry [12] was theoretically studied [13,14] and experimentally verified [15]. The particles of this geometric shape may be considered as the solid immersion lenses and natural Mie scatterers [1,2,7].

Typically, both the lateral size and curvature radius of a PH are the fraction of laser wavelength. Moreover, a PH does not have the curved side lobes [15,16]. In particular, these properties distinguish PHs and the Airy-like accelerating curved beam family [17]. Importantly, the generation of a PH is significantly easier in contrast to the formation of Airy beams. The localized light beam near the shadow surface of Janus particle bends [13] due to the constructive interference of waves passing through the particle parts with different phase velocities [13]. To date, the photonic hook phenomenon is widely studied in the scientific literature.

The influence of both axial and side illuminations with adjustable area size of spherical $BaTiO_3$ particles with Mie parameter of approximately $q = 2\pi r/\lambda \approx 42\pi$ (where $\lambda$ is illuminating wavelength and $r$ is particle radius) on optical image contrast were considered in [18]. It was shown that when the axis of the spatially limited illuminating beam is shifted toward the edge of a spherical particle, this leads to the deflection of PNJ optical axis due to the curvature of the particle surface; the PNJ deflects toward the principal symmetry axis of sphere.

A multifocal curved beam produced under on off-axis incidence of a Gaussian beam on $SiO_2$ microsphere with Mie parameter $q \sim 144\pi$, (i.e. under the geometrical-optics approximation) was considered in [19]. By means of numerical simulations, the possibility of controlling the focusing area curvature near the particle shadow hemisphere was demonstrated similar to light focusing by a cylindrical lens with decentered aperture under the normal illumination [20]. The cause of light focus bending is large spherical aberrations arising by adjusting the relative position between an off-axis Gaussian beam and a boundary of spherical particle [21]. Note that the particle sizes studied in [18,19] go beyond the range of PNJ effect existence [2,7].

Photonic hook emerged from a dielectric cylinder with embedded small glass cuboid was discussed in [22]. It was shown that by changing the incident illumination angle a curved photonic jet can be observed. This effect was understood as the interference phenomenon, which takes place within the caustic. By means of numerical simulations, two photonic hooks were observed for a specially designed five-layer dielectric micro-cylinder [23]. Analogously, twin photonic hooks generated by twin-ellipse symmetric micro-cylinders were considered in [24].

In [15] the authors noted that the observation of the PH phenomenon for different types of asymmetric mesoscale objects seems to be prospective topics for further research. Recently, for the generation of PHs the dielectric cylindrical particle with symmetry-broken material composed of solid inorganic and flexible polymer half-cylinders were proposed [25] and numerically studied in [26].

As mentioned in [15], the generation of PH using traditional elements of macro-optics appears to be a challenging task. Nevertheless, the generation of curved focus with ultrathin metalenses in the visible band under oblique illumination was recently reported in [27]. In these conditions, constructive field interference gives rise to a curved and asymmetric optical focal area, which authors [27] attributed to the photonic hook phenomenon. At the same time, this is not the case of our study because the lens dimensions do not satisfy the mesoscale condition formulated above (the Mie parameter for the lens under consideration was about $q \sim 38\pi$) and the beam curvature is several times greater than the wavelength. Moreover, the spatial structure of the side lobes in the field intensity distribution was similar to the Airy-beams distribution, whereas it is completely absent in PH-like structure.

It is worth noting, in numerous studies of photonic jets phenomenon the spherical or cylindrical particles are often employed because of their wide abundance. However, when considering optically asymmetric Janus particles, the hemispherical or semi-cylindrical particle shape [25,26] is difficult to manufacture. Here, rectangle-shaped particles are more attractive for possible PH tailoring.

Another point to be mentioned is that in all previously considered studies the microparticles were placed in the environment (air) without any stable suspension. Meanwhile, in the practical photonic applications with dielectric microparticles, their placement on a special substrate is usually required. Unfortunately, to the best of our knowledge the effect of particle fixation on a substrate on the ability to produce a PH is not yet considered. At the same time, it is known that the placement of a cubic particle with a broken symmetry on a dielectric substrate may lead to the PH spoiling because such substrate generates its own scattered field component, which interferes with the photonic hook field [28,29].

In this paper we present the results of our detailed study of the PH parameters produced by inhomogeneous (Janus) dielectric microbars placed onto an optically transparent substrate and exposed to an optical wave. New criteria of PH quality for complex flux curvature characterization is introduced and discussed. The optimal parameters of the Janus bar particle are determined for the generation of extremely curved photonic flux.

## 2. NUMERICAL SIMULATIONS METHODIC

The physical subject of the problem considered is the analysis of the near field spatial structure of light wave scattering by an inhomogeneous dielectric microparticle located on an optically transparent substrate. In mathematical formulations, this implies the solution to the diffraction problem of optical radiation in a structurally inhomogeneous medium with scattering inclusions of mesoscale dimensions, when it becomes important to take into account near-field effects due to interference of scattered and diffracted waves on a multi-contrast structure. To date, according to our information, there are no analytical solutions or quantitative estimates for such light scattering problems. Thus, the numerical simulation method is only applicable for the solution of this complicated task.

The main idea of curved photonic flux generation by means of an index-contrast Janus particle is elucidated in Figs. 1a, b. As was shown earlier [12, 13], for PH formation the optical wave should acquire asymmetric phase $\Delta\varphi$ during propagation through a particle. This can be achieved in two ways either by means of a particle with shape asymmetry $\Delta L$ (Fig. 1a) or by a particle with built-in refractive index (RI) asymmetry $\Delta n$ (Fig. 1b). Thus, desired phase modulations can be acquired using the gradient distribution of microparticle RI, e.g., by merging two materials with different optical properties $n_1$ and $n_2$ [30], as in Fig. 1b.

The near-field spatial structure upon light wave diffraction on a composite Janus particle was simulated based on the numerical solution to the wave equation for the electromagnetic field using the Finite Elements Method (FEM) implemented by the commercial software package COMSOL Multiphysics (version 4.4 for Linux). The 2D geometry of the computational domain in Cartesian coordinates $xy$ was used, and along the $z$-axis the particle was considered infinite (see, Fig. 2). The conditions of ideal field matching in the form of perfectly absorbing layers (PML) were set on the left and right boundaries of the computation domain in the direction of radiation incidence, which minimizes spurious wave reflections from the boundaries.

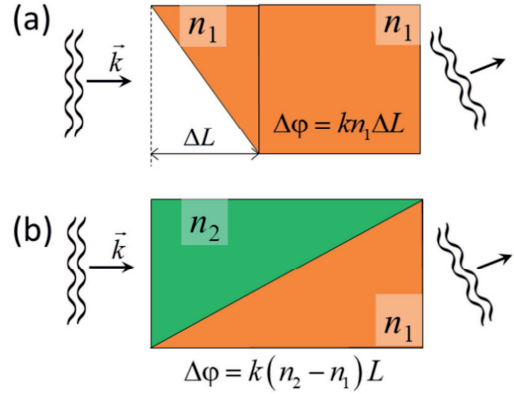

Fig.1. Principle of curved photonic flux generation by a dielectric microbar with the base *L*: (a) via shape asymmetry, (b) via refractive index asymmetry (Janus particle).

At the upper and lower boundaries of the domain, scattering conditions were employed. The numerical solution accuracy of equations was governed by an adaptive trigonal computational grid with nodes concentration in the regions of sharp gradients of medium dielectric permittivity (particle edges). The maximum spatial grid step was chosen as $\lambda/30$ inside the diffraction structure and $\lambda/15$ in the environment.

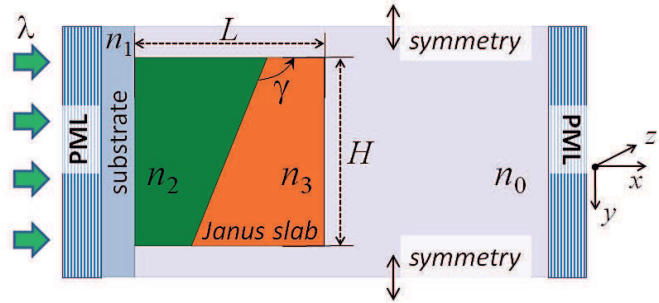

Fig. 2. 2D COMSOL model for PH generation in a Janus slab on a substrate ($n_1$) with RI contrasting parts ($n_2$ and $n_3$) illuminated by a plane wave ($\lambda$).

In the simulations, a Janus particle is represented by an orthogonal bar with geometric dimensions $H \times L$ located on a semi-infinite substrate and divided diagonally into two parts with different optical properties. The size and relative position of these parts can vary within certain limits as defined by the inclination angle of the secant diagonal. The microparticle and the substrate are considered to be made of a non-absorbing optical glass with different RIs: $n_1$, $n_2$, and $n_3$, respectively (Fig. 2). The entire diffraction assembly is illuminated orthogonally from the side face by a monochromatic optical radiation with wavelength $\lambda$ and is suspended in the environment with refractive index $n_0$.

To simulate a substrate extended along the *y*-axis providing the absence of edge diffraction, the two-step calculations of optical fields are used. At the first step, the Janus particle is removed from the geometric model and the net background optical field is calculated using the Floquet – Bloch periodicity conditions at transverse boundaries. At the second step, the microparticle is added again, and background field scattering is calculated by the composite photonic structure employing open boundary conditions.

PH formation near the shadow edge of a Janus bar exposed to optical radiation with $\lambda$ = 500 nm is illustrated in Fig. 3a, where the distribution of the normalized optical field intensity $I$ is plotted. Hereafter, for the characterizing the degree of curvature of the localized photonic flux we use several parameters shown in this figure, namely, the tilt angles of the left $\alpha_1$ and right $\alpha_2$ PH arms relative to the direction of radiation incidence, the total PH bend angle $\alpha_h = \pi - (\alpha_1 + \alpha_2)$ as well as hook height increment $h$ and the subtense $L_h$ of curved photonic flux. The slope angle calculations was carried out by a step-wise determination of the transverse coordinate $y_m$ of field intensity maximum in the PH region along each hook arm followed by the linear fitting of resulting dependences (see Fig. 3b).

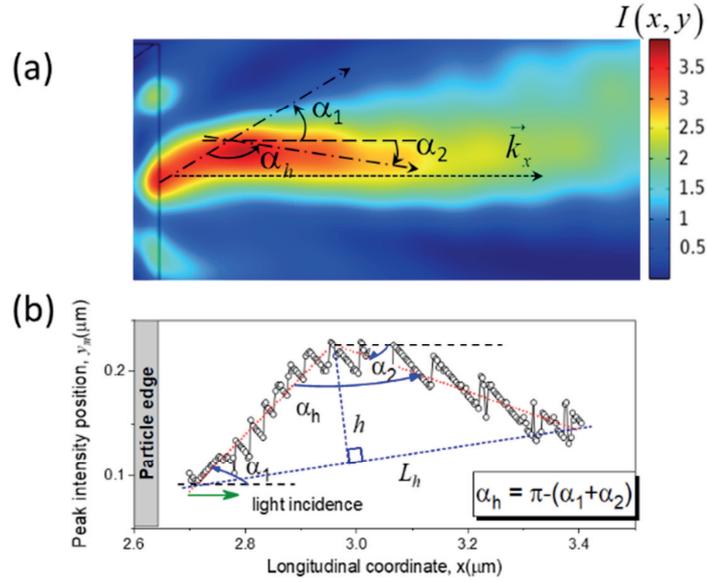

Fig. 3. (a, b) Photonic hook parameters definition: (a) normalized intensity profile $I(x, y)$ near a bar; (b) lateral position $y_m$ of hook intensity maximum versus longitudinal coordinate $x$. Here, $h$ is hook height increment, $L_h$ is subtense length.

## 3. SIMULATION RESULTS AND DISCUSSION

Consider the physical causes of the near-field optical flux bending during wave diffraction at an optically anisotropic dielectric object. To this end, it is instructive to examine the energy flows structure near a Janus particle. In Figs. 4(a, b) the spatial distribution of the time-averaged transverse component of the Poynting vector $\mathbf{S} = (c/8\pi)\mathbf{E} \times \mathbf{H}$ is shown ($\mathbf{E}$ and $\mathbf{H}$ are the electric and magnetic vectors, respectively, $c$ is the speed of light) for composite Janus bars with different RI ratios $n_2 : n_3$ of the parts. Janus particle has the dimensions $L \times H = 1.6 \times 1.5$ μm$^2$ (3.2×3 $\lambda^2$) and is placed in air ($n_0 = 1$).

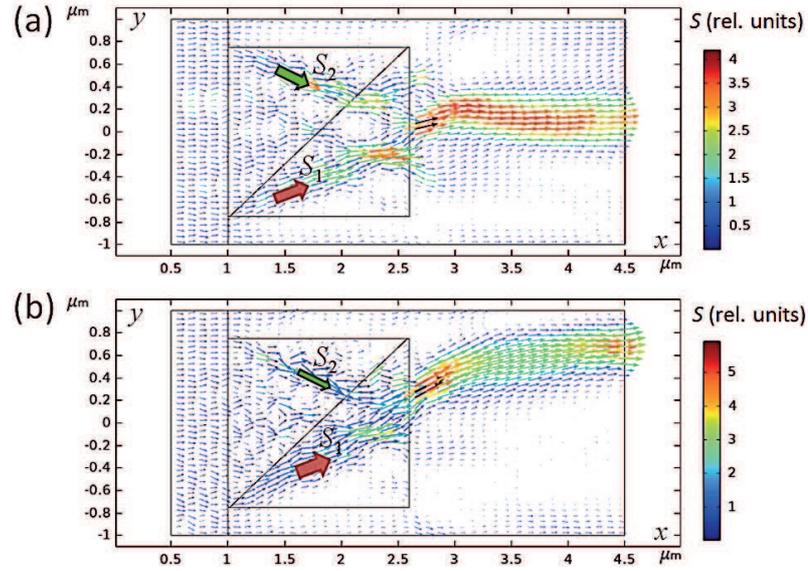

Fig. 4. (a, b) Poynting vector $S$ spatial distribution near Janus bars on substrate ($n_1$=1.5) in air ($n_0$=1) with different optical contrast (a) $n_2:n_3$=1.65:1.5 ($S_1 \approx S_2$) and (b) $n_2:n_3$=1.8:1.5 ($S_1 \gg S_2$).

As seen, the curved photonic jet is formed mainly by two most intensive optical fluxes (marked $S_1$ and $S_2$ in the figures) emerged in the upper and lower parts of the particle. Due to wave diffraction on rectangular facets, these fluxes are always directed at an angle to each other and their superposition forms a "leaky" external optical field in the form of a tilted photonic jet [13-15]. Obviously, in an optically homogeneous rectangular bar, the energy fluxes have equal intensities and the resulting photonic jet is directed along the direction of radiation incidence ($\alpha_1 = \alpha_2 = 0$). The optical contrast between the halves of a Janus particle leads to flows imbalance and energy redistribution in favor of the flux passing through the lower RI part (lower in the figures) due to the action of the general laws of light refraction at medium interface (Snell's law). As a result, the external photonic jet first acquires a "refractive" slope to the exit face of the particle ($\alpha_1 \neq 0$), and then bends ($\alpha_2 \neq 0$) due to interference of the fields in two counter-flows. Moreover, depending on the ratio of Poynting vector fluxes strength, the photon hook can substantially change its bending angle $\alpha_h$.

The dependence of PH bending angles on the refractive index $n_2$ of the upper part of Janus particle is shown in Figs. 5(a, b) for different incident wave polarizations. It is clearly seen that in both cases there is the specific range of optical contrasts, $1.01 < n_2:n_3 < 1.12$, when photonic flux bending is most pronounced (parameter $(\pi - \alpha_h)$ takes the maximum value). Maximal PH bending angle is $\alpha_h \approx 130°$ for **E**-field polarization across the plane of parts interface (*p*-polarization) and $\alpha_h \approx 150°$ for *s*-polarized wave. The latter value seems to be close to the value $\alpha_h = 148°$ obtained experimentally in the case of PH formation by a polymethylpentene particle in the THz wavelength range [15]. Here, a shape-asymmetric but optically-homogeneous particle composed by a triangular prism adjacent to the front side of a cuboid particle can be treated as a spatially symmetric but optically inhomogeneous Janus particle with rectangular faces.

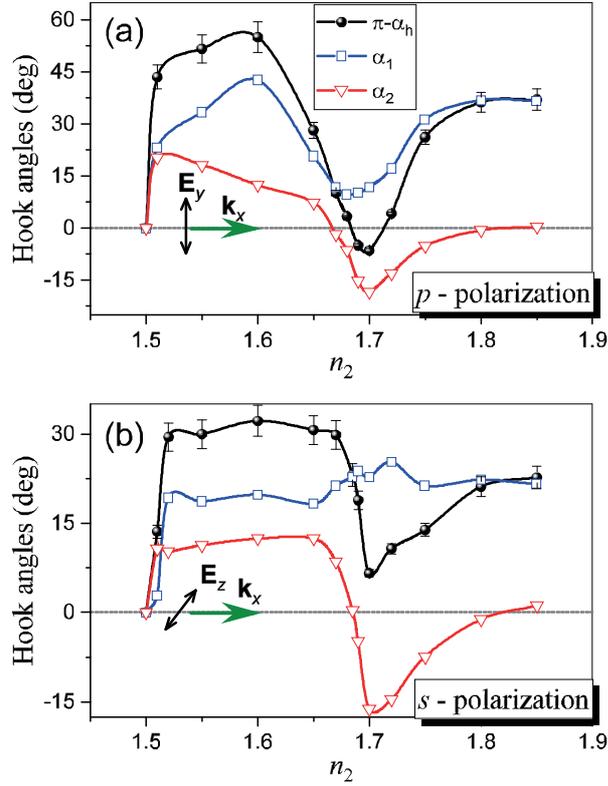

Fig. 5. (a,b) Hook angles produced by Janus bars in air (dimensions 1.6×1.5 μm$^2$) with different RI contrast for (a) *p*-polarized and (b) *s*-polarized optical wave illumination.

Beyond the range of $n_2$-values where the maximum bending of the photonic hook is observed ($n_2 > 1.68$), the angle of hook inclination $\alpha_2$ decreases due to the weakening of the energy flux $S_2$ directed from the upper most optically dense particle part (Fig. 4b). As a result, PH not only straightens out ($\alpha_2 \approx 0$) by retaining its refractive slope to the exit face of the bar ($\alpha_1 \neq 0$), but can also bend in the opposite direction (see values $\alpha_h > \pi$ in Fig. 4a). Moreover, because of the destructive interference of principal optical fluxes outside the Janus particle, PH experiences fragmentation in the longitudinal direction. With further increase in the optical contrast of bar parts, PH bend completely disappears ($\alpha_2 = 0$), and photonic hook transforms into a tilted and sufficiently elongated PNJ (~ 4λ).

Worth noting, recently considered in [26] cylindrical Janus particles with mesoscale dimensions produce a photon hook with significantly less curvature. Therefore, the authors of this work report less than 20° maximum angle of PH bending ($\delta \equiv \pi - \alpha_h$) which was achieved using Janus micro-cylinders with the optical contrast between halves $n_2/n_3 = 1.075$. This value is significantly smaller than maximal PH bend angle obtained by rectangular bars with similar characteristics in this work. At the same time, the accuracy of optical contrast for micro-cylindrical Janus particles, which is necessary to obtain the maximum bending angle, is significantly higher than for particle with an orthogonal shape. Moreover, the effect of field polarization on PH characteristics produced by cylindrical particle was left beyond the scope of Refs. [24,26].

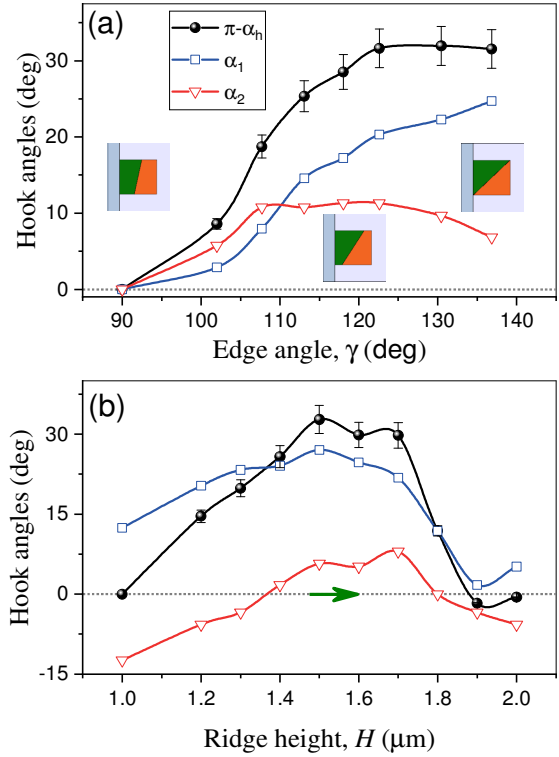

Fig. 6. Hook angles from Janus bars in the dependence on (a) edge angle and (b) bar height for *s*-polarized wave.

Hook bending angles for different geometry of the microbar are shown in Figs. 6(a, b). As expected from the geometrical-optics considerations, the increase of inclination angle $\gamma$ of the secant plane dividing Janus particle into two parts causes the increase of wave refraction angle at the interface between the parts thus introducing an imbalance between the light energy fluxes. Mostly, this affects the exit angle $\alpha_1$ of the photonic beam outside the particle whereas bending angle of the right PH arm $\alpha_2$ is influenced to a lesser extent. The highest photon hook bending corresponding to the largest $\delta$ parameter value in Fig. 6a is achieved in the case of a bar divided exactly along its diagonal (for the geometry considered, $\gamma = 136.8°$).

Fig. 6b illustrates the optimum relationship between bar geometric dimensions when PHs are formed with maximum bending between the arms. It turns out that the best in this sense are the bars with close to cubic cross-section. In this case, a decrease or conversely an increase in the aspect ratio of the bar edges leads to a hook straightening, when $\alpha_h \to \pi$. Worthwhile noting, for other polarization state of input wave (*p*-polarization) the considered dependencies are similar to those presented above and differ only in the hook angles magnitude.

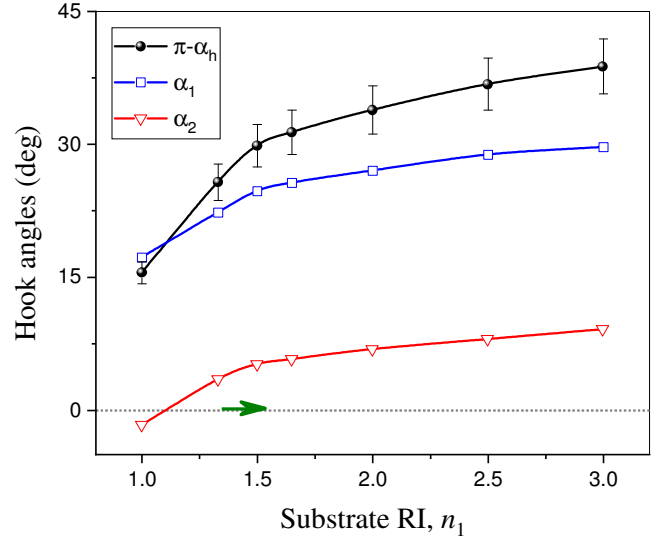

Fig. 7. Hook angles from Janus bars placed on top different substrates (*s*-polarization). Green arrow depicts wave propagation direction.

The influence of substrate RI $n_1$ on hook bending is illustrated in Fig. 7. Here, the Janus bar contrast is set as $n_2 : n_3 = 1.65:1.5$. As seen, the increase of substrate refractive index with respect to the Janus particle almost equally influences the bending angles of the left and right PH arms, specifically, leading to their increase. As a result, the net PH bending angle also increases, reaching values of $\alpha_h \sim 140°$.

One of the unique characteristics of the photonic hook is the curvature radius of principal photonic flux, which can reach subwavelength scales [12,13,15,16]. A quantitative measure of PH curvature can be the height increment $h$ defined in Fig. 3b as the distance between the center of the hook arc and the center of the subtense. From the geometrical considerations it is clear that the greater is the value of height increment $h$ and the smaller is its length $L_h$ the greater will be the curvature of the elementary light beamlet, i.e., its transversal acceleration when moving along a curved path in the analogy to the self-accelerated Airy-beams [17]. The positions of the start and end points of the chord connecting photon hook arms correspond to the coordinates of 1/*e* field intensity decrease along the hook trajectory. Note that in [24,26] the curvature of the photon hook itself was not studied, and a single value of bending angle is not enough to fully characterize the basic properties of the curved photonic flux.

Fig. 8 shows the dependences of height increment $h$ and length $L_h$ (in optical wavelengths) on the Janus particle refractive index $n_2$. Qualitatively, these two dependencies are similar to each other. The maximum value of $h$ is observed near $n_2 = 1.52$, where the subtense length is of the order of one wavelength. A further $n_2$ increase leads to a monotonic $h$ decrease up to zero value at $n_2 \approx 1.7$ when the curvature of the photonic hook is minimal (see Fig. 5a). Moreover, at $n_2 = 1.72$, the height increment $h$ becomes negative, which corresponds to PH bending inversion. These trends allow one to tailor (implement the dynamic control on) the spatial shape of the photonic hook from concave to convex beamlet by changing the refractive index $n_2$ of the Janus bar material.

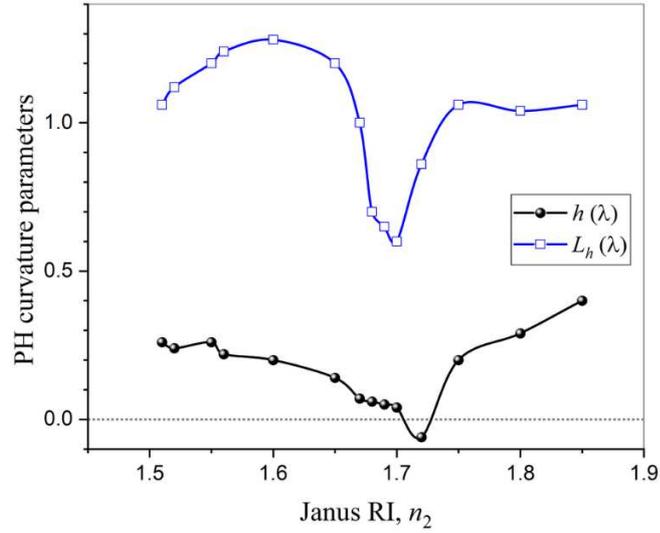

Fig. 8. PH curvature parameters versus Janus bar index $n_2$.

For complex characterizing the photon hook curvature we propose a combined hook quality criterion in the following form: $Q = I_{max} h/L_h$, where $I_{max}$ is the maximal PH field intensity. The dependence of this parameter on bar RI contrast $n_2/n_3$ is shown in Fig. 9a. The field intensity distributions corresponding to the extreme points in this dependence (b-d) are shown to the right in Figs. 9b-d.

It is clearly seen that the dependence for $Q$-criterion provides a pathway for the PH curvature manipulations. Actually, for dielectric Janus bar shown in Fig. 9b the sharp hook curvature changes are observed near particle exit surface, but already at a distance of about 1.2 µm (2.4 λ) from the shadow bar surface PH shape markedly straightens. For bar particle corresponding to Fig. 9c, the shape of the localized field becomes concave at the same distance measured from the shadow surface. The curved beamlet shown in Fig. 9d possesses the greatest curvature and field intensity $I_{max}$ resulting in the maximal value of PH quality factor $Q$.

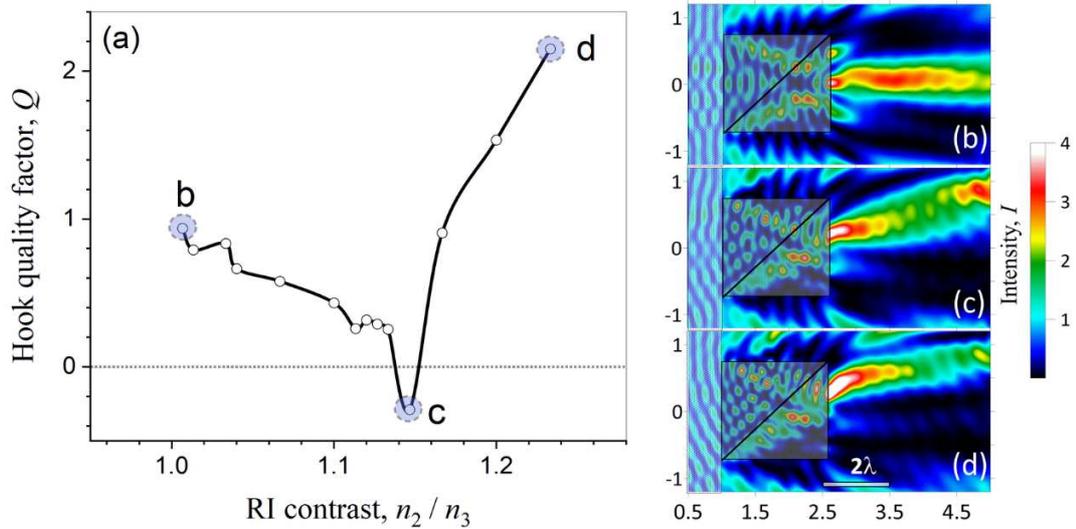

Fig. 9. (a) PH quality factor for different Janus bar RI contrast; (b-c) PH intensity plots for extreme points b-d.

## 4. CONCLUSIONS

In conclusion, we studied in the detail the principles of the formation of a new type of curved light beams — the photon hook in the scattering of a light wave by a mesoscale dielectric Janus particle. A rectangular glass bar located on a transparent dielectric substrate and consisting of two parts of the same volume with different refractive indices was proposed as such a Janus particle. The formation of photonic hooks in this case is due to the interference of the energy fluxes of light waves that diffract on different parts of the microparticles and acquire different phase incursions not because of the asymmetry of the geometric shape of the scatterer, but because of the anisotropy of its optical properties. The criteria of the photonic hook curvature quality is introduced. We show the spatial shape and curvature radius of a photonic hook can be quite simply tailored by varying the microstructure of the inhomogeneous mother particle. In this case, the maximal hook bending arises in a certain range of optical contrast of the Janus bar halves $1.01 < n_2/n_3 < 1.12$ and increases with increasing refractive index of the dielectric substrate.

## 5. FUNDING

Y.G. was supported by the Ministry of Science and Higher Education of the Russian Federation. I.M. and O.M. were partially supported by the Russian Foundation for Basic Research (Grant No. 20-57-S52001) and the work partially was carried out within the framework of the Tomsk Polytechnic University Competitiveness Enhancement Program, Russia.

**Author Contributions**

I.M. and O.M. conceived the idea and supervised the project; Y.G. performed numerical simulations; I.M., O.M., Y.G. prepared the paper draft and interpreted the results; All authors contributed to the manuscript.